\documentclass[fleqn,usenatbib]{mnras}
\usepackage{newtxtext,newtxmath}
\usepackage{float}
\usepackage{threeparttable}
\usepackage{ulem}
\usepackage{comment}
\usepackage{indentfirst}
\usepackage{here}
\usepackage{afterpage}
\usepackage{multicol}

\usepackage[T1]{fontenc}

\DeclareRobustCommand{\VAN}[3]{#2}
\let\VANthebibliography\thebibliography
\def\thebibliography{\DeclareRobustCommand{\VAN}[3]{##3}\VANthebibliography}

\usepackage[dvipdfmx]{graphicx} 
\usepackage{amsmath}	

\makeatletter
\def\myfigure{\vbox\bgroup\centering\def\@captype{figure}}
\def\mytable{\vbox\bgroup\centering\def\@captype{table}}
\def\endmyfigure{\egroup}
\def\endmytable{\egroup}
\makeatother

\title[Discovery of NEI plasma associated with the NPS/Loop I]{Discovery of non-equilibrium ionization plasma associated with the North Polar Spur and Loop I}

\author[Yamamoto et al.]{Marino Yamamoto,$^{1}$\thanks{E-mail: maricat@akane.waseda.jp}
Jun Kataoka,$^{1}$
Yoshiaki Sofue$^{2}$
\\
$^{1}$Faculty of Science and Engineering, Waseda University, 3-4-1, Okubo, Shinjuku, Tokyo 169-8555, Japan\\
$^{2}$Institute of Astronomy, The University of Tokyo, 2-21-2, Osawa, Mitaka-shi, Tokyo 181-0015, Japan}

\date{Accepted 2022 March 1. Received 2022 March 1; in original form 2021 December 6}

\pubyear{2021}

\begin{document}
\label{firstpage}
\pagerange{\pageref{firstpage}--\pageref{lastpage}}
\maketitle

\begin{abstract}
We investigated the detailed plasma condition of the North Polar Spur (NPS)/Loop I using archival $Suzaku$ data. In previous research collisional ionization equilibrium (CIE) have been assumed for X-ray plasma state, but we also assume non-equilibrium ionization (NEI) to check the plasma condition in more detail. 
We found that most of the plasma in the NPS/Loop I favors the state of NEI, and has the density-weighted ionization timescale of $n_e t\sim10^{11-12}$ s cm$^{-3}$ and the electron number density $n_e\sim$ a few $\times$ 10$^{-3}$ cm$^{-3}$. 
The plasma shock age, $t$, or the time elapsed after the shock front passed through the plasma, is estimated to be on the order of a few $\rm{Myr}$ for the NPS/Loop I, which puts a strict lower limit to the age of the whole NPS/Loop I structure. We found that NEI results in significantly higher temperature and lower emission measure than those currently derived under CIE assumption.
The electron temperature under NEI is estimated to be as high as 0.5~keV toward the brightest X-ray NPS ridge at $\Delta\theta=-20^\circ$, which decreases to 0.3 keV at $-10^\circ$, and again increases to $\sim 0.6$ keV towards the outer edge of Loop I at $\Delta\theta\sim0^\circ$, about twice the currently estimated temperatures.
Here, $\Delta \theta$ is the angular distance from the outer edge of Loop I.
We discuss the implication of introducing NEI for the research in plasma states in astrophysical phenomena.
\end{abstract}



\section{Introduction} \label{sec:intro}
Large structures in the direction of the Galactic Center (GC) have been discovered by observations at various wavelengths. An excess of microwaves, called \textit{WMAP} haze \citep{2004ApJ...614..186F}, has been identified by \textit{WMAP}, and this has also been confirmed by  \textit{Planck} observations \citep{2013A&A...554A.139P}. Numerous Galactic spurs and loops have been observed by radio continuum surveys (e.g., \citet{1971A&A....14..252B,1982A&AS...47....1H}), and more recently by \textit{WMAP} \citep{2015MNRAS.452..656V}, of which Loop I is the largest structure, spanning $\sim120^\circ$ on the sky. Similarly, in X-rays, Loop I has been confirmed by the  \textit{ROSAT} all-sky survey (RASS) at 0.75 keV \citep{1995ApJ...454..643S}, and a particularly bright structure in its northeastern part is called the North Polar Spur (NPS). Although RASS did not find corresponding features in the southern part of the GC, an all-sky survey of \textit{eROSITA} launched in 2019 \citep{2021A&A...647A...1P} discovered relatively weak, but giant bubble-like structures, \textit{eROSITA} bubbles, with a north-south symmetry and extending over 80$^{\circ}$ against the GC. $\gamma$-ray observations by the Large Area Telescope (LAT;\citet{2009ApJ...697.1071A}) on \textit{Fermi} have also revealed the existence of bubbles structure extending $50^\circ$ above and below the GC, which is called the Fermi bubble \citep{2010ApJ...724.1044S}.  It has been argued that the symmetric north-south bubbles observed by \textit{Fermi} and \textit{eROSITA} may be a remnant of the GC explosion in the past \citep{2000ApJ...540..224S, 2021MNRAS.506.2170S}.

It is still debated whether NPS/Loop I is a nearby (100$-$200 pc) structure associated with recent  supernova remnants (SNRs) or a distant structure formed by the outflow from the GC over several Myr ago.
When NPS/Loop I was first discovered, the prevailing theory was that it is a nearby SNR in the vicinity of the Sun because of its north-south asymmetry. Moreover, the visual alignment of the stellar/dust polarization with the synchrotron emission in the radio band also supports this idea \citep{1967MNRAS.137..157B,1971A&A....14..252B,1995A&A...294L..25E}. After the discovery of the Fermi bubbles, however, the competing idea that it is a remnant of   the GC explosion \citep{1977A&A....60..327S,2000ApJ...540..224S, 2003ApJ...582..246B} has come into the spotlight. The X-ray emission from NPS/Loop I  is well represented by a thin thermal emission at $kT$ $\simeq$ 0.3 keV under the process of ionization equilibrium by \textit{Suzaku} and other observations \citep{2013ApJ...779...57K,2018ApJ...862...88A,2020ApJ...904...54L,2021ApJ...908...14K}. This is slightly higher than the typical Galactic halo (GH) temperature of 0.2 keV\citep{2009PASJ...61..805Y}; thus it can be interpreted as shock-heated GH gas. In this context, the shock velocity after the GC explosion was estimated as $v_{sh}\sim300 \; \rm{km\:s^{-1}}$ , with the total energy released being $10^{55-56}\rm{erg}$ and the explosion occurring over ~10 Myr ago \citep{2018Galax...6...27K}. Nevertheless, the distance to NPS/Loop I is still debated based on the most recent measurements of radio polarization and the X-ray absorption measurements due to the interstellar medium (ISM)  \citep{2015MNRAS.447.3824S,2021arXiv210614267P}.

In either formation scenario, whether due to the GC explosion or a nearby SNR, the material ejected by the explosion collides with the surrounding ISM to form a collisionless shock wave. In the case of a young SNR, it is generally observed that the plasma is in a  state of non-equilibrium ionization (NEI) \citep{2002A&A...381.1039W, 2003ApJ...589..827B}, where the ionization temperature $T_z$ is not equal to either the temperature of the electrons  ($T_e$) or that of protons ($T_p$). Such a state of non-equilibrium develops when the heating of the plasma occurs in a very short timescale compared to the relaxation time. The degree of ionization of the plasma at $t$ is characterized by a density-weighted timescale $n_e t$, which is expressed as a product of electron density $n_e$ and shock age $t$, where $n_et\sim10^{12} \rm{s\:cm^{-3}}$ is required to reach an equilibrium for the plasma \citep{2010ApJ...718..583S}.
Note that if the electron density, $n_e$ is measured from the emission measure of the observed X-ray spectrum, then the age of the SNR plasma since it was heated, $t$ which is called shock age, can be estimated from the above density-weighted timescale. Note that shock age is the time since the plasma was heated by the shock wave, where density-weighted timescale $n_e t=0$ in front of the shock wave ($t=0$) \citep{2001ApJ...548..820B}. Similarly, a timescale in which the NPS/Loop I plasma is heated may be estimated from $n_e t$ of the X-ray emitting plasma. However, all previous studies have assumed that plasma in the NPS/Loop I is in CIE, but such an assumption may not hold, particularly in low-density ($n_e$ $\ll$ 1 cm$^{-3}$) plasma such as the GH gas.

In this study, we revisited the archival data of the $Suzaku$ X-ray Imaging Spectrometer (XIS : \cite{2007PASJ...59S..23K})  to examine the degree of ionization equilibrium in the NPS/Loop I region 
to estimate their origin.  The remainder of this paper is organized as follows: Section \ref{sec:data} describes the analysis region and criteria for data reduction when preparing the data to be used for the analysis. In Section \ref{sec:result}, we describe the analysis method used to extract the spectra of X-ray diffuse emission and its analysis results. We discuss the physical origin of the NPS/Loop I by summarizing the results obtained from the analysis in Section \ref{sec:discus}. We also discuss future prospects. All statistical errors in the text and tables are $1\sigma$, unless otherwise stated.

\section{Observation and data reduction} \label{sec:data}
\subsection{Archival Data and Analysis Region} 
We used the \textit{Suzaku} XIS archive of X-ray data for our analysis. The \textit{Suzaku} satellite \citep{2007PASJ...59S...1M} is equipped with four focal-plane X-ray CCD cameras (XIS), a X-ray microcalorimeter (XRS : \citet{2007PASJ...59S..77K}) at the focal points of five reflectors, XRT, and a non-imaging hard X-ray detector (HXD : \citet{2007PASJ...59S..35T}) that measures observations above 10~keV. Among the four XISs, XIS0, 2 and 3 are front-illuminated (FI) and XIS1 is back-illuminated (BI), covering the energy range of 0.2$-$12 keV. However, XIS2 was damaged in November 2006 due to contamination by a leaked charge; therefore, data from the other three XISs were used in this analysis. We did not analyze the HXD data because the anticipated emissions from the NPS in this band are too weak compared to both the instrumental and cosmic X-ray background (CXB) \citep{2013ApJ...779...57K}.
The \textit{Suzaku} is particularly suitable for studying diffuse emission, as relevant for this work, because of its high energy resolution at low energies and its low background. 

\begin{figure*}
	\includegraphics[width=160mm]{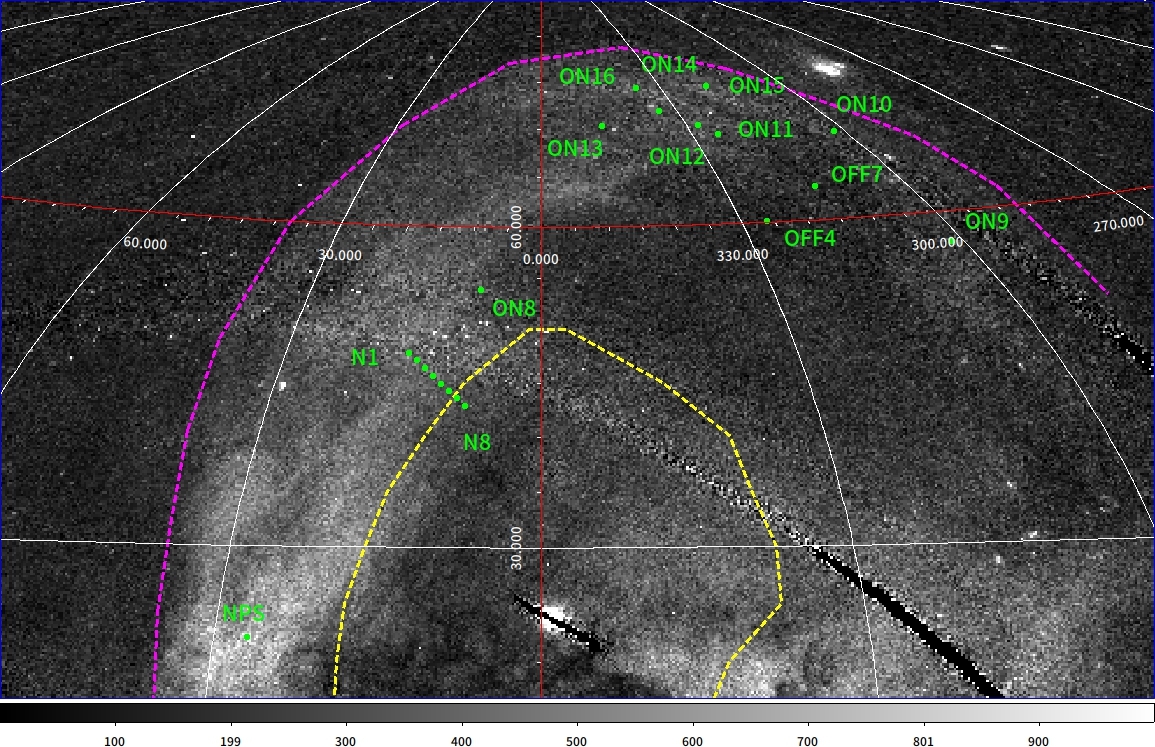}
    \caption{Observation region by \textit{Suzaku} shown on a 0.75 keV image of the $ROSAT$. The region corresponds to the northern part of the Fermi bubbles. This image is displayed on a linear scale, and the color bar at the bottom is $10^6$ counts s$^{-1}$. Numbers written in white are galactic coordinates for longitude and latitude. The green circle corresponds to Suzaku's XIS field of view (17.8 arcmin). The yellow dashed line is the boundary of the bubble proposed in \citet{2010ApJ...724.1044S}, and the magenta dashed line is the NPS/Loop I structure's outer edge (see, Sec.\ref{sec:3.2} for more details).}
    \label{fig:region}
\end{figure*}

In this analysis, we used the region of bright emission in RASS as measured in the 0.75 keV (R34) band around the northern bubble, such as the NPS/Loop I structure. For the low-latitude NPS/loop I region ($b<50^{\circ}$), we used data from the brightest region \citep{2008PASJ...60S..95M} and the eight pointing regions across the edge of the northeast bubble were observed as part of the AO7 program \citep{2013ApJ...779...57K}. In addition, in the high-latitude NPS/Loop I ($b>50^{\circ}$), the regions analyzed in a previous study \citep{2018ApJ...862...88A} were selected for analysis. The regions and information used in this analysis are shown in Figure \ref{fig:region} and Table \ref{tab:data}.

\begin{table*}
    \caption{\textit{Suzaku} observation}
	\centering
	\label{tab:data}
	\begin{tabular}{ccccccccc} 
		\hline
		Name & Obs ID & Start Time & Stop Time & R.A.& Decl. & \textit{l} & \textit{b} & Exposure \\
         &  &  & (UT) & (deg) $\;^{a}$ & (deg)$\;^{b}$ & (deg) $\;^{c}$ & (deg) $\;^{d}$ & (ks) $\;^{e}$ \\
		\hline
		\multicolumn{9}{c}{low-latitude NPS/Loop I region ($b<50^{\circ}$)}\\
        \hline
        N1 & 507006010 & 2012 Aug 8 10:23 & 2012 Aug 8 23:03 & 233.401 & 9.076 & 15.480 & 47.714 & 19.2\\
        N2 & 507005010 & 2012 Aug 7 23:41 & 2012 Aug 8 10:22 & 233.623 & 8.079 & 14.388 & 47.011 & 15.1\\
        N3 & 507004010 & 2012 Aug 7 10:31 & 2012 Aug 7 23:40 & 233.834 & 7.087 & 13.321 & 46.308 & 19.3\\
        N4 & 507003010 & 2012 Aug 6 23:20 & 2012 Aug 7 10:30 & 234.034 & 6.098 & 12.280 & 45.606 & 18.1\\
        N5 & 507001010 & 2012 Aug 5 23:04 & 2012 Aug 6 09:33 & 234.250 & 5.090 & 11.255 & 44.871 & 16.4\\
        N6 & 507002010 & 2012 Aug 6 09:34 & 2012 Aug 6 23:18 & 234.405 & 4.131 & 10.263 & 44.204 & 20.9\\
        N7 & 507007010 & 2012 Aug 8 23:06 & 2012 Aug 9 10:20 & 234.551 & 3.174 & 9.291 & 43.537 & 18.0\\
        N8 & 507008010 & 2012 Aug 9 10:21 & 2012 Aug 9 23:53 & 234.713 & 2.200 & 8.334 & 42.838 & 20.9\\
        NPS & 100038010 & 2005 Oct 3 11:29 & 2005 Oct 4 10:50 & 260.588 & 4.757 & 26.840 & 21.960 & 39.5\\
        \hline
        \multicolumn{9}{c}{high-latitude NPS/Loop I region ($b>50^{\circ}$)}\\
        \hline
        ON-8 & 802038010 & 2007 Jul 14 17:09 & 2007 Jul 15 13:50 & 225.629 & 8.293 & 7.819 & 53.735 & 28.2\\
        ON-9 & 805041010 & 2011 Jan 14 11:37 & 2011 Jan 17 04:34 & 192.151 & $-5.791$ & 301.636 & 57.074 & 76.2\\
        ON-10 & 802039010 & 2007 Jun 17 20:11 & 2007 Jun 18 22:10 & 192.513 & 5.456 & 301.998 & 68.325 & 33.4\\
        ON-11 & 509062010 & 2014 Dec 20 00:12 & 2014 Dec 20 12:36 & 200.607 & 7.382 & 324.764 & 68.930 & 13.8\\
        ON-12 & 509059010 & 2014 Dec 23 18:49 & 2014 Dec 24 05:16 & 201.171 & 8.665 & 327.544 & 69.932 & 18.3\\
        ON-13 & 702053010 & 2008 Jan 7 07:23 & 2008 Jan 8 16:19 & 206.913 & 12.350 & 347.389 & 70.203 & 42.6\\
        ON-14 & 804048010 & 2010 Jan 23 01:08 & 2010 Jan 24 14:00 & 202.357 & 11.020 & 333.767 & 71.580 & 42.0\\
        ON-15 & 701080010 & 2006 Jun 18 05:29 & 2006 Jun 19 05:04 & 197.244 & 11.578 & 318.614 & 73.913 & 28.5\\
        ON-16 & 805074010 & 2010 Dec 17 19:04 & 2010 Dec 18 00:57 & 201.743 & 13.573 & 336.170 & 74.105 & 12.3\\
        OFF-4 & 702067010 & 2007 Jul 14 08:31 & 2007 Jul 14 17:00 & 204.195 & $-0.819$ & 326.078 & 59.999 & 12.0\\
        OFF-7 & 705011010 & 2011 Jan 17 04:38 & 2011 Jan 17 17:54 & 198.188 & 0.850 & 314.829 & 63.228 & 39.5\\
		\hline
	\end{tabular}
	{\footnotesize \\
	    Note - The names N1-N8 refer to \citet{2013ApJ...779...57K}, NPS to \citet{2008PASJ...60S..95M}, and Loop I ON and OFF to the region analyzed in \citet{2018ApJ...862...88A}. \\
	    $^{a}$ R.A. (Right ascension) of \textit{Suzaku} pointing center in J2000 equinox. \\
        $^{b}$ Decl. (Declination) of \textit{Suzaku} pointing center in J2000 equinox. \\
        $^{c}$ Galactic longitude of \textit{Suzaku} pointing center. \\
        $^{d}$ Galactic latitude of \textit{Suzaku} pointing center. \\
        $^{e}$ Exposure time of a good time interval after the data reduction described in Section \ref{sec:data}.\\}
\end{table*}

\subsection{Data Reduction of XIS}
To reduce and analyze the $Suzaku$ data, we used \textsc{heasoft} version 6.27, \textsc{xspec} version 12.11.0, \textsc{xronos} version 6.0, and calibration database (\textsc{caldb}) released on October 10, 2018.  We  followed the processes described in ‘The Suzaku Data Reduction Guide’ \footnote{ \url{https://heasarc.gsfc.nasa.gov/docs/suzaku/analysis/abc/}}. First, \textsc{xselect} was used to combine the data of different observation modes ($3\times3$, $5\times5$) to the cleaned events that have already been screened for XIS 0, 1, and 3. These criteria were used to perform data reduction based on the mkf file using \textsc{xselect} : (1) data during and up to 436 s after the South Atlantic Anomaly (SAA) are not used ; (2) data above $5^\circ$ and $20^\circ$ from the rim of the earth (ELV) at night and daytime, respectively ; (3) we selected data with a cutoff rigidity (COR) higher than 6 GeV ; and (4) remove data when the satellite is turned off by more than 1.5 arcmin from the mean pointing angle to eliminate misalignment of the satellite's attitude after the maneuver. Furthermore, calibration sources, hot pixels, and flickering pixels were excluded based on the information regarding the status of each event.

Solar wind charge exchange (SWCX) can cause contamination of the $\rm{O_{VII}}$ emission line contamination near 0.525 keV. Since SWCX is correlated with the proton flux in the solar wind, we did not use the data taken when the solar wind flux was above $4.0\times10^8 \; \rm{cm^{-2}s^{-1}}$ \citep{2009PASJ...61..805Y}. In addition, we checked the light curve around 0.5 keV and no flares occurred during the time of use.
The proton flux was from the \textsc{omniweb}\footnote{\url{https://omniweb.gsfc.nasa.gov/form/dx1.html}}. The ELV of the day-earth was set to more than $40^\circ$ in the ON 11 region because the $\rm{O_{VII}}$ emission could not be sufficiently removed.

\subsection{Extracting Spectra}
Before extracting the spectra, we created images of XIS 0, 1 and 3 in 0.4$-$10 keV. When generating images, count rate correction was performed with \textsc{xisexpmapgen} \citep{2007PASJ...59S.113I} and non-X-ray background (NXB) estimated from the night Earth observation data was removed with \textsc{xisnxbgen} \citep{2008PASJ...60S..11T}. These images were also corrected for vignetting using \textsc{xissim} \citep{2007PASJ...59S.113I}.

\textsc{ximage} was used to detect point sources in the field of view with a significance of at least $3\sigma$ for the combined XIS0 and XIS3 images in order to improve the statics. For the analysis of diffuse emissions, we created and used a region file that excludes regions with a radius of more than 2 arcmin, centered on the coordinates of the detected source in the field of view. For the regions that excluded point sources, we then created spectra of all events and NXB spectra using \textsc{xisnxbgen}. We also created redistribution matrix files (RMFs) and auxillary response files (ARFs) using xisrmfgen and xissimarfgen \citep{2007PASJ...59S.113I}. ARFs were created assuming uniform radiation from a circle of radius 20 arcmin.

\section{Results} \label{sec:result}
\subsection{Spectral Analysis} \label{subsec:3.1}
\begin{figure}
    \includegraphics[width=\columnwidth]{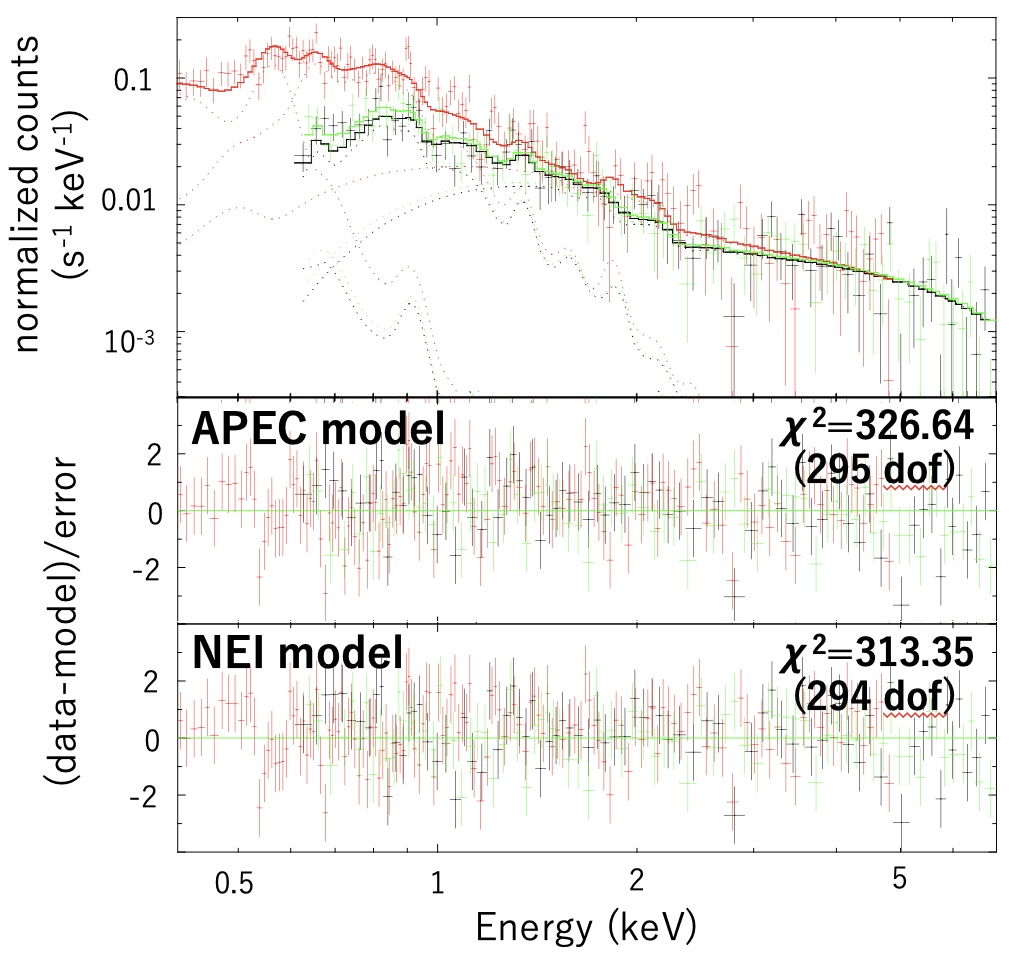}
    \caption{An example of data analysis (in N8 region). XIS spectra of the diffuse emission with the best-fit model curve in the upper column. The XIS0 data/model are shown in black, XIS1 in red, and XIS3 in green. The middle panels show the \texttt{apec} model, and the bottom panels show the \texttt{nei} model and data residual. See the text for details on fitting the models.}
    \label{fig:spectrum}
\end{figure}

We fit the spectra using \texttt{XSPEC}, with the energy channels chosen so that each bin contained at least 20 photons. The energy range used is 0.6$-$7.0 keV for XIS0 and 3, and 0.4$-$5.0 keV for XIS1 due to its sensitivity at low energy and high background at high energy because of BI. An example (N8 region) of the spectrum analysis result is shown in Figure \ref{fig:spectrum}.

Following previous works, the X-ray diffuse emission has been modeled by a combination of three emission components \citep{2009PASJ...61..805Y,2013ApJ...779...57K}; (1) unabsorbed thermal emission from local hot bubbles (LHB) and/or SWCX at a fixed $kT$ = 0.1 keV and $Z = Z_{\odot}$, (2) absorbed thermal emission related to the GH and NPS/Loop I at a fixed $Z = 0.2 Z_{\odot}$, and (3) absorbed cosmic X-ray background (CXB). The neutral absorption column density $N_{\rm{H, Gal}}$ is fixed at the full Galactic column value \footnote{\url{https://cxc.harvard.edu/toolkit/colden.jsp}} \citep{1990ARA&A..28..215D}. LHB/SWCX is represented  \texttt{apec}, which is the model of CIE, and CXB is expressed by a power law with the photon index fixed at $\Gamma = 1.41$, as determined by \citet{2002PASJ...54..327K}. The normalization of all components tied between instruments was set to free.

In addition, the contribution of unabsorbed thermal plasma with $kT \sim 0.1$ keV is proposed to account for low-energy spectra below 0.5~keV \citep{2008PASJ...60S..95M,2020ApJ...904...54L}. Therefore, we also performed the analysis by adding an absorbed 
$kT \sim 0.1$ keV thermal plasma. The results showed that the parameters for (2), namely, temperature, EM, and the density weighted timescale of the GH and NPS/Loop I components, are unchanged within the error range (see, Appendix for an example case of the spectral fitting of N8). Moreover, statistical improvement is negligible in most cases; a part of the reason may be due to the relatively low photon statistics of $Suzaku$ XIS below 0.5~keV. Therefre, in this paper, we thus followed the three-component model initially proposed by \citet{2013ApJ...779...57K}, and subsequent papers.

Especially important for this analysis is that we utilized two different models to represent the thermal plasma emission of the GH component: \texttt{apec} and \texttt{nei}, which correspond to the CIE and NEI, respectively. The fitting results for the GH component are presented in the Table\ref{tab:results}. Note that the emission measure (EM) is calculated from Eq. (1) of the \citet{2013ApJ...766...48S} by considering the ARF and assuming a circular radiation region of 20 arcmin.

\begin{table*}
    \caption{Results of fitted parameters of diffuse emission of GH in NPS/Loop I \label{tab:results}}
	\centering
	\label{tab:data}
	\begin{tabular}{ccccccccc}
		\hline
		 & \multicolumn{3}{c}{\texttt{apec} model (CIE)} & \multicolumn{4}{c}{\texttt{nei} model (NEI)} & F$-$test \\
        \cline{2-9}
        Name & $kT_{\rm{GH}}\;^{a}$ & $\rm{EM_{GH}}\;^{b}$ & $\chi^2$/dof &  $kT_{\rm{GH}}\;^{c}$ & $\rm{EM_{GH}}\;^{d}$ & $n_et_{GH}\;^{e}$ & $\chi^2$/dof & P$-$value$\;^{f}$ \\
         & (keV) & ($10^{-2}\rm{cm^{-6}pc}$) &  & (keV) & ($10^{-2}\rm{cm^{-6}pc}$) & ($\rm{10^{11} s/cm^{-3}}$) &  &\\ 
		\hline
		\multicolumn{9}{c}{low-latitude NPS/Loop I region ($b<50^{\circ}$)}\\
    \hline
    N1 & $0.307_{-0.010}^{+0.011}$ & $5.31_{-0.35}^{+0.37}$ & 293.45/280 & $0.307_{-0.010}^{+0.023}$ & $5.27_{-0.80}^{+0.32}$ & $>4.40$ & 291.36/279 & $1.58\times10^{-1}$\\
    N2 & $0.315_{-0.012}^{+0.014}$ & $5.27_{-0.39}^{+0.39}$ & 297.78/259 & $0.373_{-0.050}^{+0.050}$ & $3.66_{-0.76}^{+1.29}$ & $3.86_{-1.51}^{+5.20}$ & 294.11/258 & $7.39\times10^{-2}$\\
    N3 & $0.311_{-0.011}^{+0.012}$ & $5.61_{-0.37}^{+0.40}$ & 282.59/297 & $0.395_{-0.045}^{+0.046}$ & $3.38_{-0.59}^{+0.88}$ & $2.90_{-0.96}^{+2.12}$ & 278.05/296 & $2.87\times10^{-2}$\\
    N4 & $0.303_{-0.009}^{+0.010}$ & $5.76_{-0.39}^{+0.37}$ & 344.16/300 & $0.442_{-0.056}^{+0.034}$ & $2.69_{-0.29}^{+0.76}$ & $1.85_{-0.42}^{+1.22}$ & 336.78/299 & $1.10\times10^{-2}$\\
    N5 & $0.277_{-0.010}^{+0.010}$ & $5.65_{-0.42}^{+0.46}$ & 285.00/271 & $0.508_{-0.042}^{+0.052}$ & $1.71_{-0.23}^{+0.0.25}$ & $0.90_{-0.25}^{+0.34}$ & 269.05/270 & $8.16\times10^{-5}$\\
    N6 & $0.298_{-0.012}^{+0.014}$ & $3.85_{-0.33}^{+0.34}$ & 290.45/274 & $0.505_{-0.058}^{+0.060}$ & $1.44_{-0.22}^{+0.29}$ & $1.21_{-0.37}^{+0.57}$ & 285.01/273 & $2.32\times10^{-2}$\\
    N7 & $0.274_{-0.011}^{+0.011}$ & $5.00_{-0.39}^{+0.48}$ & 274.49/249 & $0.556_{-0.058}^{+0.161}$ & $1.27_{-0.32}^{+0.25}$ & $0.60_{-0.38}^{+0.33}$ & 264.21/248 & $8.80\times10^{-3}$\\
    N8 & $0.268_{-0.009}^{+0.010}$ & $4.69_{-0.38}^{+0.40}$ & 326.64/295 & $0.605_{-0.074}^{+0.115}$ & $1.05_{-0.18}^{+0.17}$ & $0.49_{-0.21}^{+0.23}$ & 313.35/294 & $4.80\times10^{-4}$\\
    NPS & $0.291_{-0.003}^{+0.003}$ & $15.2_{-0.29}^{+0.29}$ & 958.22/654 & $0.459_{-0.008}^{+0.010}$ & $6.26_{-0.14}^{+0.26}$ & $1.53_{-0.10}^{+0.13}$ & 914.41/653 & $3.28\times10^{-8}$\\
    \hline
    \multicolumn{9}{c}{high-latitude NPS/Loop I region ($b>50^{\circ}$)}\\
    \hline
    ON-8 & $0.320_{-0.011}^{+0.013}$ & $2.98_{-0.20}^{+0.19}$ & 454.20/417 & $0.415_{-0.053}^{+0.038}$ & $1.80_{-0.24}^{+0.50}$ & $2.74_{-0.66}^{+2.02}$ & 448.29/416 & $1.97\times10^{-2}$ \\
    ON-9 & $0.282_{-0.008}^{+0.010}$ & $3.67_{-0.25}^{+0.25}$ & 447.44/455 & $0.536_{-0.047}^{+0.047}$ & $1.10_{-0.12}^{+0.17}$ & $0.84_{-0.21}^{+0.31}$ & 431.28/454 & $4.42\times10^{-5}$\\
    ON-10 & $0.272_{-0.010}^{+0.010}$ & $2.50_{-0.19}^{+0.21}$ & 447.66/407 & $0.658_{-0.085}^{+0.497}$ & $0.58_{-0.20}^{+0.11}$ & $0.52_{-0.39}^{+0.26}$ & 428.58/406 & $2.64\times10^{-5}$\\
    ON-11 & $0.342_{-0.020}^{+0.027}$ & $2.65_{-0.29}^{+0.28}$ & 247.95/148 & $0.547_{-0.031}^{+0.046}$ & $1.27_{-0.13}^{+0.14}$ & $1.43_{-0.33}^{+0.26}$ & 233.29/147 &$2.81\times10^{-3}$\\
    ON-12 & $0.347_{-0.021}^{+0.026}$ & $2.92_{-0.30}^{+0.31}$ & 233.7/224 & $0.577_{-0.017}^{+0.042}$ & $1.35_{-0.12}^{+0.11}$ & $1.20_{-0.24}^{+0.27}$ & 201.82/223 & $1.11\times10^{-8}$\\
    ON-13 & $0.340_{-0.013}^{+0.017}$ & $2.37_{-0.18}^{+0.16}$ & 498.43/445 & $0.499_{-0.028}^{+0.033}$ & $1.28_{-0.11}^{+0.12}$ & $1.86_{-0.32}^{+0.39}$ & 477.34/444 & $1.19\times10^{-5}$\\
    ON-14 & $0.327_{-0.012}^{+0.012}$ & $3.03_{-0.18}^{+0.21}$ & 664.23/568 & $0.537_{-0.025}^{+0.027}$ & $1.33_{-0.09}^{+0.10}$ & $1.33_{-0.18}^{+0.21}$ & 403.5/567 & $2.27\times10^{-63}$\\
    ON-15 & $0.259_{-0.006}^{+0.006}$ & $5.34_{-0.27}^{+0.31}$ & 394.93/360 & $0.459_{-0.033}^{+0.057}$ & $1.43_{-0.20}^{+0.16}$ & $0.45_{-0.14}^{+0.20}$ & 403.5/359 & - \\
    ON-16 & $0.256_{-0.012}^{+0.012}$ & $4.47_{-0.45}^{+0.56}$ & 188.18/179 & $0.540_{-0.095}^{+0.121}$ & $1.00_{-0.18}^{+0.31}$ & $0.53_{-0.21}^{+0.43}$ & 185.91/178 & $1.42\times10^{-1}$\\
    OFF-4 & $0.396_{-0.075}^{+0.208}$ & $0.68_{-0.25}^{+0.25}$ & 153.15/154 & $0.582_{-0.138}^{+0.146}$ & $0.39_{-0.10}^{+0.17}$ & $1.52_{-0.82}^{+2.45}$ & 150.44/153 & $9.89\times10^{-2}$\\
    OFF-7 & $0.252_{-0.017}^{+0.020}$ & $2.00_{-0.32}^{+0.40}$ & 201.88/216 & $0.748_{-0.217}^{+17.420}$ & $0.33_{-0.11}^{+0.13}$ & $0.29_{-0.25}^{+0.39}$ & 197.05/215 & $2.27\times10^{-2}$\\
	\hline
    \end{tabular}
    {\footnotesize \\
    Note - We fixed the metal abundance of the GH in the NPS/Loop I to $Z=0.2Z_\odot$. The absorption column densities for the GH in the NPS/Loop I and CXB components were fixed to the Galactic values given by Dickey \& Lockman (1990).\\
    $^{a}$ Temperature of thermal plasma of the GH under the assumption of CIE (\texttt{apec} model) in NPS/Loop I.\\
    $^{b}$ Emission measure of thermal plasma of the GH under the assumption of CIE (\texttt{apec} model) in NPS/Loop I.\\
    $^{c}$ Temperature of thermal plasma of the GH under the assumption of NEI (\texttt{nei} model) in NPS/Loop I.\\
    $^{d}$ Emission measure of thermal plasma of the GH under the assumption of NEI (\texttt{nei} model) in NPS/Loop I.\\
    $^{e}$ Density-weighted timescale of thermal plasma of the GH under the assumption of NEI (\texttt{nei} model) in NPS/Loop I. It reaches ionization equilibrium at approximately $n_et\simeq10^{12}$.
    $^{f}$ The probability that the chi-square distributions of the \texttt{apec} and \texttt{nei} models coincide, estimated using the F-test.
    }
\end{table*}

\begin{figure*}
    \begin{center}
    \includegraphics[width=110mm]{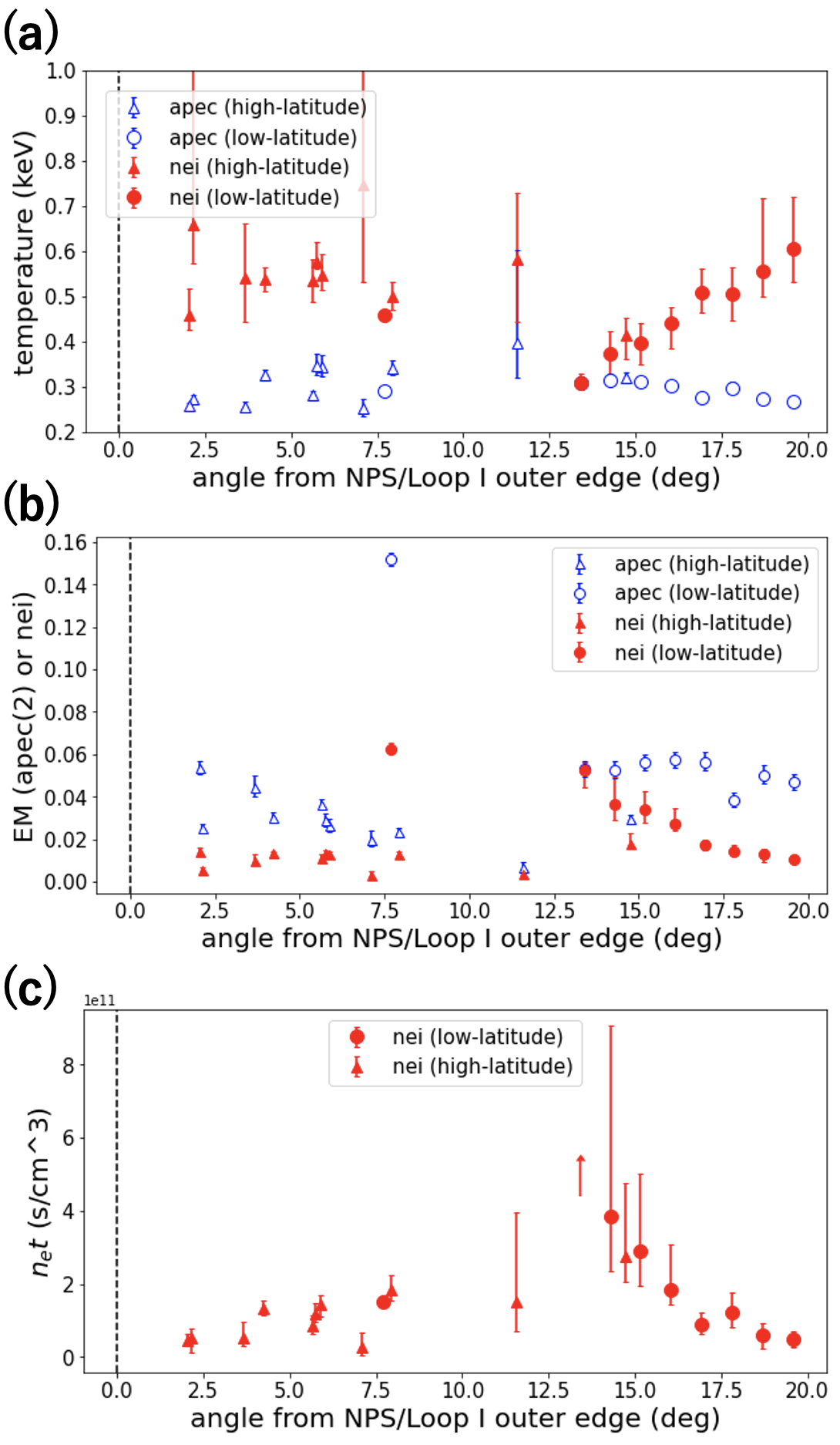}
    \caption{Comparison of fitting results of the GH components in the NPS/Loop I region using \texttt{apec} and \texttt{nei} models. These are plotted against the angular distance from the inner edge of the NPS/Loop I defined in Table \ref{tab:coord} and Fig.\ref{fig:region} represented by magenta dashed line.}(a) Electron temperature, $kT$. (b) Emission measure (EM). (c) Density-weighted timescale $n_e t$. 
    \label{fig:results}
    \end{center}
\end{figure*}

\subsection{Spatial variation of $kT$, $EM$ and $n_et$}\label{sec:3.2}
Figure \ref{fig:results} shows the variation of $kT$, EM, and density-weighted timescale determined from the analysis as measured in the regions in NPS/Loop I. Note that the horizontal axis shows the separation angle from the outer edge of the NPS/Loop I structure defined by the coordinate in Table \ref{tab:coord} and the magenta dashed line in Fig.\ref{fig:region}. These coordinates of the outer edge of this NPS/Loop I are defined to be consistent with the eRosita bubble observed by the allsky survey of eRosita \citep{2020Natur.588..227P}. The angle increases from the outer to the inner parts of the northern bubble; thus, the GC direction corresponds to a positive value. Furthermore, the temperature estimated in this fitting is the electron temperature, which will be denoted as $kT$ in this paper.

For all regions, the EM of the GH component was larger in the \texttt{apec} model than in the \texttt{nei} model. Focusing on the continuous observation of low-latitude NPS/Loop I region (N1-N8) in particular, the variation of the EM with the angle from the edge is more obvious in the \texttt{nei} model than in the \texttt{apec} model.

We find that the electron temperature steeply decreases from 0.5~keV at  $\Delta \theta\sim -20^\circ$ to 0.3 keV at $-10^\circ$, and then increases to $\sim 0.6$ keV towards the outer edge at $\Delta \theta\sim 0^\circ$, where $\Delta \theta$ is the angular distance from the outer edge of Loop I toward inside.

Figure \ref{fig:results} also indicates that regions with plasma closer to the CIE state show larger density-weighted timescale and lower temperature.
This means that when the electron density is high, the density-weighted timescale will also be large because $n_e$ is large if the shock age is comparable. In the brightest spot of the NPS, $n_e t\simeq 1.5\times10^{11} \; \rm{s\:cm^{-3}}$; thus, it has not yet reached the CIE state. For the high-latitude NPS/Loop I region, there is no significant position dependence, but assuming the \texttt{nei} model, the plasma is in the overall NEI state and the temperature of the plasma is approximately $kT\simeq0.6 \rm{keV}$.

\begin{figure}
\begin{center}
    \includegraphics[width=80mm]{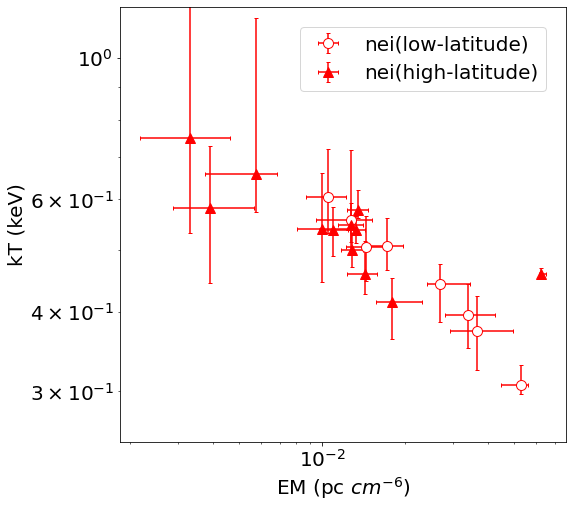}
    \caption{Correlation between EM and temperature $kT$ for GH plasmas in NPS/Loop I under the assumption of NEI state. In this figure, the $circles$ represents low-latitude, the $filled triangles$ represent high-latitude.}
    \label{fig:corr}
    \end{center}
\end{figure}

Note that, the EM and electron temperature $kT$ of the GH component of the NPS/Loop I plasma have a good negative correlation under the assumption of the \texttt{nei} model. This correlation between the EM and electron temperature is shown in Fig.\ref{fig:corr}. The correlation coefficient between the EM and plasma temperature is $-0.858$ assuming the NEI state, which is significantly larger than the negative correlation of $-0.484$ obtained from when assuming the CIE state.

\subsection{F-test} \label{subsec:3.3}
According to Table \ref{tab:results}, the \texttt{nei} model provides better fitting results than \texttt{apec} in most regions, although the results of \texttt{apec} are acceptable in terms of the values of the reduced chi-square ($\chi^2\rm{/dof}$).
To evaluate the validity of introducing the \texttt{nei} model, we conducted an F-test. The F-test compares two chi-square distributions and examines the relationship between them under the null hypothesis that the two distributions are equally distributed. If the null hypothesis is correct, then $F_\chi$ 
\begin{equation}
    \label{eq:f-distribution}
    F_\chi = \frac{\chi^2_1/\nu_1}{\chi^2_2/\nu_2}
\end{equation}
follows an F distribution, where $\chi^2_1, \chi^2_2$ are the chi-square of the \texttt{apec} and \texttt{nei} models with degrees of freedom $\nu_1$ and $\nu_2$. The smaller $F_\chi$ is, the more consistent the two chi-square distributions are, and the probability that the models are consistent is expressed by 
\begin{equation}
    \label{eq:p-value}
    P_F(F; \nu_1, \nu_2)=\int^\infty_Fp_f(f; \nu_1, \nu_2) df
\end{equation}
where $p_f(f; \nu_1, \nu_2)$ is the probability density function expressed using the $\Gamma$ function \citep{Bevington:1305448}. The p-values for the application of the \texttt{apec} and \texttt{nei} models in the analysis are shown in Table \ref{tab:results}. In most regions, the P-value is below 0.05, so the null hypothesis is rejected considering a 95\% confidence interval and statistical significance between two models is guaranteed.

It is, however, noted that the \texttt{apec} model is also able to reproduce the spectrum effectively, so CIE cannot be completely rejected from statistical perspective. Since the \texttt{apec} model represents the CIE, we can say that the density-weighted parameter in the \texttt{apec} model is $n_et\sim10^{12-13} \; \rm{s\:cm^{-3}}$ or higher.
Therefore, the \texttt{nei} contains another free parameter, which is density-weighted parameter, when fitting the data, thus reduced chi-squared values tend to be smaller than corresponding \texttt{apec} model. 

For example, the plasma in N1, which is the farthest region from the GC in the low-latitude NPS (see, Fig. \ref{fig:region}), has almost reached the CIE, so the density-weighted timescale $n_et$ in fitting assuming the NEI can only give a lower limit. Thus, if the plasma has reached ionization equilibrium, the \texttt{apec} and \texttt{nei} models lead to the same spectrum. In fact, in the N1 region, the fitting results of the \texttt{apec} and \texttt{nei} models are consistent within the error (see, Table \ref{fig:results}). This result indicates that the NEI includes the CIE, which is physically natural.

\section{Discussion and conclusion} \label{sec:discus}
\subsection{Corroboration of explosion in GC} \label{subsec:4.1}

\begin{table}
	\centering
	\caption{Definition of the outer coordinates of the NPS/Loop I arc.}
	\label{tab:coord}
	\begin{tabular}{c} 
		\hline
		($l,b$ [deg])\\
		\hline
        (33.1, 9.9)\\
        (34.5, 16.2)\\
        (35.5, 24.0)\\
        (36.3, 31.7)\\
        (37.6, 40.1)\\
        (38.4, 48.6)\\
        (37.6, 59.9)\\
        (28.9, 70.0)\\
        (9.6, 77.0)\\
        (333.4, 78.5)\\
        (308.6, 75.6)\\
        (295.4, 71.0)\\
        (288.5, 67.2)\\
        (286.3, 61.8)\\
        (287.6, 51.3)\\
		\hline
	\end{tabular}
	{\footnotesize\\
	Note - $l$ is Galactic longitude, and $b$ is Galactic latitude in Galactic coordinate. \\}
\end{table}

\begin{figure*}
    \begin{center}
    \includegraphics[width=110mm]{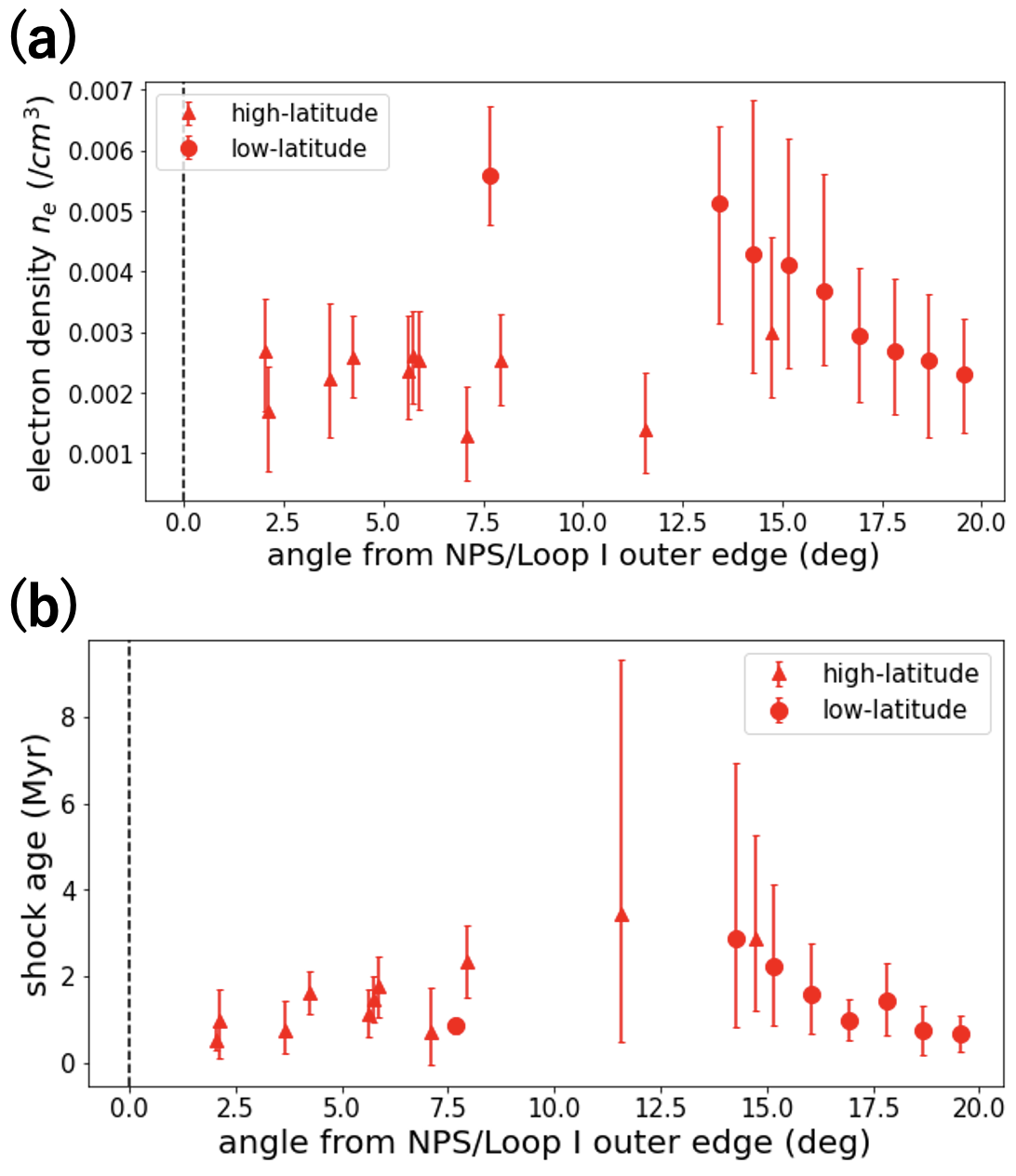}
    \caption{Spatial distribution of (a)electron density $n_e$ , and (b)shock age $t$ , assuming that the plasma is in the NEI state and explosion was around the GC. These are plotted against the angular distance from the inner edge of the NPS/Loop I defined in Table \ref{tab:coord} and Fig.\ref{fig:region} represented by magenta dashed line.}
    \label{fig:calc}
    \end{center}
\end{figure*}

Table \ref{tab:results} shows that the density-weighted timescale $n_et$ of NPS/Loop I is of the order of $n_et\simeq10^{11} \rm{s\:cm^{-3}}$, which indicates that the plasma is close to CIE, but still in the state of NEI. On the other hand, the electron density $n_e$ of the plasma is estimated to be as small as $n_e\sim3\times10^{-3} \rm{cm^{-3}}$ \citep{2015ApJ...807...77K}.
In our case, the measured range of $n_e$ was in the range of (2$-$5)$\times$10$^{-3}$cm$^{-3}$ assuming the scale length of the structure is 2~kpc (see Eq.(1) in \citet{2013ApJ...766...48S} for the conversion of EM to electron number density $n_e$), and thus is consistent with \citet{2015ApJ...807...77K}. The electron density in each region is calculated on the assumption that the explosion occurs near the GC, as shown in Fig.\ref{fig:calc}. Also, the energy of the explosion can be estimated as $E \simeq n_g \times kT \times V \sim 10^{54-55}\rm{erg}$, where $n_g$ $\simeq$ $n_e$ is the gas number density, and $V$ is the total volume of the compressed NPS/Loop I gas.

In this paper, we assumed that the absorbed hot thermal plasma exists in the NPS and is regarded as a single component; however, there are also other studies in which the absorbed $kT \sim 0.1$ keV thermal plasma exists in addition to $kT \sim 0.3$ keV plasma \citep{2008PASJ...60S..95M,2020ApJ...904...54L}. They also consider the absorbed thermal plasma of $kT \sim 0.1$ keV to be a cool component of the NPS. In fact, the EM of the unabsorbed $kT \sim 0.1$ keV thermal plasma, which we consider to be the LHB/SWCX component in our analysis, is significantly larger than the average EM of the LHB before excluding the SWCX contribution \citep{2014ApJ...791L..14S}.
However, the details of the origin and the amount of absorption of $kT \sim 0.1$ keV are still under debate. 

For example, \citet{2015ApJ...808...22H} analyzed a large set of $XMM$-$Newton$ and $Suzaku$ shadowing observations to discriminate the relative contributions of LHB and SWCX and discussed the possible uncertainties in the estimated temperature and brightness of the Galactic halo. These satellites carry CCD cameras with high spectral resolutions, allowing important line emission features to determine the plasma temperature (e.g., O VII and O VIII) to be resolved. Unfortunately, these CCD cameras are  only usable above $\sim$0.4 keV 
energy band; thus, it is difficult to determine the spectral properties of $kT$ $\simeq$ 0.1 keV plasma. Similarly, \citet{2017ApJ...834...33L} revisited the RASS data to produce the "cleaned" maps of only the LHB. They incorporated the most recent observations from a sound rocket mission, $DXL$, to quantify and characterize the contribution of SWCX to RASS data as measured in the 1/4 keV band. However, an accurate estimation of the SWCX is still difficult, owing to low-spectral-resolution proportional counters on board $ROSAT$. Moreover, the contribution of SWCX varies largely both in the time and space. They concluded that the total energy enclosed in the LHB may be more than 15 times smaller that estimated without removing the SWCX contribution. Together with an unknown amount of absorption as described above, an accurate measurement of the LHB is yet far from being complete and requires further clarifications.

Thus, we attempted to fit the data by assuming another extreme case, in which $kT$ $\sim$ 0.1 keV plasma is fully absorbed with $N_{\rm H}$ = $N_{\rm Gal}$; however, the results were unchanged for all the other remaining parameters, namely,  normalization, the temperature of $kT \sim 0.3$ keV plasma, and the normalization of the CXB, within a one sigma statistical uncertainty. These results suggested that the absorbed thermal plasma of the NPS was in the NEI state, irrespective of the origin or amount of interstellar absorption of the plasma at $kT \sim 0.1$ keV. Also, if there exists an absorbed thermal plasma of $kT \sim 0.1$ keV exist in the NPS, as suggested by \citet{2020ApJ...904...54L}, this cool component may dominate the total energy stored in the NPS, which further supports our hypothesis that the NPS is created by a huge past explosion of the GC.

Importantly, such a large density-weighted timescale $n_e t$ 
with such a low electron number density $n_e$ requires a large shock age $t$. In fact, assuming the \texttt{nei} model, the shock age is calculated from the EM and density-weighted timescale to be around $t\sim\rm{Myr}$. 

Fig.\ref{fig:calc} shows the calculation of the shock age $t$ based on the electron density $n_e$ and density-weighted timescale $n_e t$ in each region. 
This time can be interpreted as the elapsed time after the shock front arrived and heated the plasma in the corresponding position of the GH. 
Therefore, this time must be shorter than the age of the present shock front forming the NPS.
This is consistent with the estimated age of the bubble produced by a nuclear outburst occurred in the GC about 10~Myr ago \citep{1977A&A....60..327S,2000ApJ...540..224S}. 


Conversely, if the NPS/Loop I structures are associated with a nearby SNR, these structures may be located approximately $\sim$100 pc apart from the Sun. Even in such a case, we can calculate the electron density and shock age in the same way, but we obtain $n_e\simeq2\times10^{-2}\rm{cm^{-3}}$ for the electron density assuming the depth of the gas is 50 pc, and $t\simeq300~\rm{kyr}$ for the shock age. Thus, the plasma density is much lower than the typical ISM density of $n_e\sim1\rm{cm^{-3}}$. For example, a shell of the Cygnus loop, which is an SNR located at the distance of 540 pc from the sun, has an electron density of $n_e\simeq1.0\; \rm{cm^{-3}}$ \citep{2006A&A...447..937S}, and the shell of the superbubble in the OB association has an electron density of $n_e\simeq3\; \rm{cm^{-3}}$ \citep{1986PASJ...38..697T}. Thus, the density of NPS/Loop I is exceptionally thin compared to those of the shell in the SNRs near the solar system. In addition, the shock age of the NPS/Loop I is too long compared to that of typical Galactic SNRs. Furthermore, the energy of the structure, assuming an SNR, is $E$ $\simeq$ $n_g$$\times$$kT$$\times$$V$ $\sim10^{52}\rm{erg}$, which is almost equivalent to the energy emitted by a few dozen typical type-II SNRs, and is thus not plausible. Of course, the energy $E$ can vary largely with the volume $V$, which depends sensitively on the distance to the structure. However, if the NPS/Loop I was really a nearby SNR close to the sun, it is unusual that the corresponding optical emission, such as bright H$\alpha$ filaments as seen in Cygnus Loop \citep{1998ApJS..118..541L}, is hardly seen in the case of NPS/Loop I.

\subsection{Plasma State in NPS/Loop I}
In this context, the comparison of timescales to reach equilibrium in the X-ray emitting plasma provides additional support for the \texttt{nei} state in the NPS/Loop I structure. It is believed that this plasma is swept up by the shock and then heated to emit X-ray. When the plasma is heated by the shock wave, the proton temperature is initially $m_p/m_e$ $\simeq$ 1836 times higher than that of electrons, where $m_p$ and $m_e$ are the masses of protons and electrons, respectively. Over time, the kinetic energy of the protons is transferred to the electrons, moving closer to thermal equilibrium. By adopting the equations in \citet{1962pfig.book.....S} normalized to typical values in the NPS/Loop I region, we obtain the relaxation time, which is equal to the time of equipartition, as  
\begin{equation}
    \label{eq:time-ep}
    t_{ep}\sim1.1\times10^6\left(\frac{n_e}{3\times10^{-3}\rm{cm^{-3}}}\right)^{-1}\left(\frac{kT_e}{0.3~\rm{keV}}\right)^{3/2} \rm{yr}
\end{equation}
where $T_e$ is the electron temperature. For comparison, the self-collision time due to collisions between protons and electrons in each group is expressed by 
\begin{equation}
    \label{eq:time-pp}
    t_{pp}\sim2.6\times10^5\left(\frac{n_e}{3\times10^{-3}\rm{cm^{-3}}}\right)^{-1}\left(\frac{kT_e}{0.3~\rm{keV}}\right)^{3/2} \rm{yr}
\end{equation}
\begin{equation}
    \label{eq:time-ee}
    t_{ee}\sim620\left(\frac{n_e}{3\times10^{-3}\rm{cm^{-3}}}\right)^{-1}\left(\frac{kT_e}{0.3~\rm{keV}}\right)^{3/2} \rm{yr}
\end{equation}
where $t_{pp}$ is the proton-proton  and $t_{ee}$ is the electron-electron self-collision time, and after this time, these particles attain a Maxwell distribution.

On the other hand, instantaneous heating via the shock wave  breaks the ionization state from its initial equilibrium state, resulting in the NEI state. In order that such NEI plasma reaches an approximate CIE, $n_et\simeq10^{12}$ cm$^{-3}$s is necessary. Hence a typical timescale for the transition 
to the CIE state is given by 
\begin{equation}
    \label{eq:time_CIE}
    t_{\rm{CIE}}\sim10^7\left(\frac{n_e}{3\times10^{-3}\rm{cm^{-3}}}\right)^{-1} \rm{yr}
\end{equation}
.

As calculated in Section \ref{subsec:4.1} and Eq. \ref{eq:time-ep}, \ref{eq:time-pp}, and \ref{eq:time-ee}, it can be said that the electron temperature $T_e$ and proton temperature $T_p$ are close to equilibrium, that is, $T_e\sim T_p$ if NPS/Loop I is a Myr-scale event. However, Eq. \ref{eq:time_CIE} shows that the ionization state of the plasma, represented by the temperature $T_z$, does not reach equilibrium, and is most likely in the NEI state. This is consistent with the analysis in Section \ref{subsec:3.3}, where the CIE represents the overall features of the observed X-ray spectra, while the NEI provides a better representation of data in terms of the results of reduced chi-squared and F-test. It is important to note that statistically, CIE cannot be completely rejected. If the plasma state is in CIE, the density-weighted timescale is $n_et\simeq10^{12}$ $\rm{s \;cm^{-3}}$, and the shock age is close to Eq. \ref{eq:time_CIE}.

\subsection{Time Evolution of the Plasma}
Continuous observations of the low-latitude NPS/Loop I region across the bubble edge and analysis assuming the \texttt{nei} model show that the electron temperature is cooler, brighter, and closer to the CIE toward the outside of the GC (see, red circle plot in Fig. \ref{fig:results}). Such a gradient may be caused by the time evolution of the shock wave. The temperature is lower toward the outer side of the GC because of the slower shock wave, but brighter because a higher amount of ISM has been swept up. Furthermore, these tendency appears to be similar to what was anticipated from the radial profile of the density and temperature described by the self-similar solution under the assumption that the Shock front is mostly outside of the GC. However, further careful simulation or modeling is necessary. HaloSat observations show a similar temperature gradient within the NPS \citep{2020ApJ...904...54L}. In addition, the density-weighted timescale gradually increased toward the outer regions of the  NPS, reflecting a gradual increase in the plasma density $n_e$.

It is not clear whether the plasma temperature and the velocity of the shock wave directly connected, but if we consider the high-latitude region close to the outer edge of NPS/Loop I as the shock front, we can roughly estimate the velocity of the shock wave as in \citet{2013ApJ...779...57K} assuming the \texttt{nei} model.
Because a typical GH temperature is 0.2 keV and the speed of sound is approximately $C_s\sim200 \rm{km\:s^{-1}}$, and according to Rankine-Hugoniot equation, the observed electron temperature of the high-latitude NPS/Loop I $kT\sim0.5-0.6 \rm{keV}$ corresponds to a shock velocity with a Mach number $M\geq2.3-2.7$. Hence the velocity of the shock wave in NPS/Loop I is estimated to be $v\geq460-530\;\rm{km\:s^{-1}}$. In this context, \citet{2016ApJ...829....9M} derived a similar range of the $v=490^{+230}_{-77}\; \rm{km\;s^{-1}}$ expansion velocity using the  $\rm{O_{VII}}$ and $\rm{O_{VIII}}$ emission lines, although the 
paper assumed a slightly different model that contains an underlying $kT$ $\simeq$ 0.2 keV halo emission that co-exists with the shocked component described above.

Although gradual changes in $kT$, EM, and $n_e t$ are seen in the continuous observation of low-latitude NPS with distance from the NPS/Loop I inner edge, such a sequential trend does $not$ seem to smoothly connect with the parameters in the high-latitude NPS/Loop I, as shown in Fig. \ref{fig:results}. Meanwhile, the plasma in the Loop I is also well described by NEI with an density-weighted timescale of about $n_e~t\sim10^{11}\rm{s\:cm^{-3}}$ and an electron temperature of $kT\sim 0.5-0.6 \rm{keV}$, which is significantly higher than typical GH temperatures. This fact may suggest that the NPS/Loop I structure have resulted from multiple explosions. However, the observations in the high-latitude NPS/Loop I are not spatially continuous and sparse, so the cause of the lack of continuity is most likely undersampling.Sensitive all-sky surveys with soft X-rays, such as eROSITA, will provide us with new results to address these problems. Additional systematic surveys are necessary to discuss the past activity of the galaxy in detail.

\section*{Acknowledgements}
We thank an anonymous referee for many useful comments to improve the manuscript. This work was supported by JST ERATO Grant Number JPMJER2102 and JSPS KAKENHI grant No.JP20K20923, Japan.

\section*{Data Availability}

 The data from the XIS detector by $Suzaku$ is available from heasarc archieve (\url{https://heasarc.gsfc.nasa.gov/cgi-bin/W3Browse/w3browse.pl}). \textsc{caldb} was used for calibration data, and the \textsc{heasoft} suite of tools was used for data analysis.



\bibliographystyle{mnras}
\bibliography{main} 




\appendix
\section{Detailed modeling of the Plasma with $kT$ $\simeq$ 0.1 keV}
As discussed in the text (Sections \ref{subsec:3.1} and \ref{subsec:4.1}), we analysed assuming a single unabsorbed $kT$ $\simeq$ 0.1 keV plasma to mimic the low-energy X-ray emission from LHB/SWCX. However, certain studies \citep{2008PASJ...60S..95M,2020ApJ...904...54L} have proposed the existence of cool component $kT$ $\simeq$ 0.1 keV of NPS, in addition to LHB/SWCX emission. The origin of such  cool component is still far from being understood. Thus, in this appendix, we thus tried to fit the same spectra with various emission models, namely (1) double NPS (CIE + CIE), in which both cool ($kT$ $\simeq$0.1 keV) and hot ($kT$ $\simeq$0.3 keV) components exist in the NPS and modeled as CIE, (2) double NPS (CIE+NEI), where the same spectrum is  modeled as CIE + NEI, and (3) absorbed LHB, where the temperature of the cool component is fixed at $kT$ $\simeq$ 0.1 keV but absorbed as the same amount with the NPS. The LHB/SWCX component in (1) and (2) ia absorbed at the galactic value of the neutral absorption column density $N_{\rm{H, Gal}}$. The example results of spectral fitting for the N8 region in the low-latitude NPS/Loop I region are summarized in Table \ref{tab:append}. Note that, irrespective of the amount of absorption assumed, the temperature of the cool component is always converged to $kT$ $\simeq$ 0.1 keV and that of the hot component is $kT$ $\simeq$ 0.6 keV, respectively. Thus, the results are consistent with that shown in  Section \ref{subsec:3.1} within 1 $\sigma$ statistical uncertainty. Therefore, we conclude that the discussion and conclusion in this study would not be affected even though the physical origin of $kT$ $\simeq$0.1 keV cool component remains uncertain.

\setcounter{table}{3}
\begin{table*}
    \caption{Results of fitted parameters of diffuse emission of GH in NPS/Loop I in N8}
	\centering
	\label{tab:append}
	\begin{tabular}{ccccccccc}
		\hline
		& & LHB & \multicolumn{2}{c}{NPS(cool component)} & \multicolumn{3}{c}{NPS(hot component)} &  \\
        \cline{3-8}
        number & model & EM & kT & EM & kT & $n_et_{GH}$ & EM & $\chi^2$/dof \\
        & & ($10^{-2}\rm{cm^{-6}pc}$) & (keV) & ($10^{-2}\rm{cm^{-6}pc}$) & (keV) & ($\rm{10^{11} s/cm^{-3}}$) & ($10^{-2}\rm{cm^{-6}pc}$) &\\ 
		\hline
        (1) & double NPS (CIE+CIE) $\rm{\;^{a}}$ & $< 3.18$ & $0.092_{-0.007}^{+0.001}$ & $28.1_{-21.9}^{+22.1}$ & $0.268_{-0.010}^{+0.011}$ & $-$ & $4.33_{-4.09}^{4.24}$ & 325.37/293\\
        (2) & double NPS (CIE+NEI) $\rm{\;^{b}}$ & $<1.56$ & $0.081_{-0.023}^{+0.013}$ & $74.4_{-34.9}^{+51.1}$ & $0.661_{-0.116}^{+0.132}$ & $3.05_{-1.23}^{+3.01}$ & $0.94_{-0.13}^{0.22}$ & 307.64/292\\
        (3) & absorbed LHB $\rm{\;^{c}}$ & $5.75_{-0.72}^{+0.55}$ & $-$ & $-$ & $0.608_{-0.067}^{+0.115}$ & $5.43_{-2.22}^{+2.32}$ & $0.981_{-0.170}^{+0.146}$ & 312.94/294\\
 	    \hline
    \end{tabular}
    {\footnotesize \\
    Note - We fixed the metal abundance of the GH in the NPS/Loop I to $Z=0.2Z_\odot$. The absorption column densities for the GH in the NPS/Loop I and CXB components were fixed to the Galactic values given by Dickey \& Lockman (1990).\\
    $\rm{^a}$ \texttt{apec + wabs*(apec + apec + powerlaw)} \\
    $\rm{^b}$ \texttt{apec + wabs*(apec + nei + powerlaw)} \\
    $\rm{^c}$ \texttt{wabs*(apec + apec + powerlaw)} \\
    }
\end{table*}



\bsp	
\label{lastpage}
\end{document}